\documentclass[12pt]{iopart}
\begin{document}

\title{Neural coding of naturalistic motion stimuli}
\author{G. D. Lewen, W. Bialek and R. R. de Ruyter van 
Steveninck\\ NEC Research Institute\\ 
4 Independence Way\\
Princeton, New Jersey 08540 \\}

\maketitle

\begin{abstract}

We study a wide field motion sensitive neuron in the
visual system of the blowfly {\em Calliphora vicina}. By
rotating the fly on a stepper motor outside in a wooded
area, and along an angular motion trajectory 
representative of natural flight, we stimulate the fly's
visual system with input that approaches the natural 
situation. The neural response is analyzed in the 
framework of information theory, using methods that are 
free from assumptions. We demonstrate that 
information about the motion trajectory increases as 
the light level increases over a natural range. This 
indicates that the fly's brain utilizes the increase in 
photon flux to extract more information from the 
photoreceptor array, suggesting that imprecision in neural signals
is dominated by photon shot noise in the physical input,
rather than by noise generated within the nervous system
itself.

\end{abstract}

\section{Introduction}  

One tried and tested way to study sensory information
processing by the brain is to stimulate the sense organ
of interest with physically appropriate stimuli and to
observe the responses of a selected part of the system
that lends itself to measurement. Within that framework
there are strong incentives, both practical and
analytical, to simplify stimuli. After all, short
lightflashes or constant tones are easier to generate
and to capture mathematically than the everchanging
complex world outside the laboratory. Fortunately, sense
organs and brains are extremely adaptive, and they
apparently function in sensible ways, even in the
artificial conditions of typical laboratory experiments.
A further reason for using simplified stimuli is that
they are presumed to elicit simple responses,
facilitating interpretation of the system's input-output
behaviour in terms of underlying mechanism. Typically,
these simple stimuli are repeated a large number of
times, the measured outputs are averaged, and this
average is defined to be the `meaningful' component of
the response. This is especially helpful in the case of
spiking neurons, where we face the embarrasment of the
action potential: Because they are an extremely
nonlinear feature of the neural response we often do not
really know how to interpret sequences of action 
potentials (Rieke \etal, 1997). One way to evade the 
question and save tractability is to work with derived 
observables, in particular with smooth functions of 
time, such as the average firing rate.

Although it certainly is useful to perform experiments
with simplified inputs, one also would 
like to know how those stimuli are processed and encoded
that an animal is likely to encounter in nature. We 
expect animals to be `designed' for those conditions, 
and it will be interesting to see to what extent the 
brain can keep up with the range and strength of these 
stimuli. In the present work  we are concerned 
primarily with the question of how noisy neural 
information processing really is. This question cannot 
be answered satisfactorily if we do not study the 
problem that the brain is designed to solve, because for
us it is hard to distinguish willful neglect on the part
of the brain in solving artificial tasks, from noisiness
of its components.

As soon as we try to characterize the behaviour of a
sensory system in response to the complex, dynamic,
nonrepeated signals presented by the natural world, we
lose many of the simplifications mentioned earlier. To
meet the challenge we must modify both our experimental
designs and our methods for analyzing the responses to
these much more complicated inputs. Recent examples of
laboratory based approaches to the problem of natural
stimulation are studies of bullfrog auditory neurons
responding to synthesized frog calls (Rieke \etal, 
1995), insect olfactory neurons responding to odour 
plumes (Vickers \etal, 2001), cat LGN cells responding 
to movies (Dan \etal, 1996, Stanley \etal, 1999), 
primate visual cortical cells during free viewing of 
natural images (Gallant \etal, 1998, Vinje and Gallant, 
2000), auditory neurons in song birds stimulated by song
and song--like signals (Theunissen and Doupe, 1998, 
Theunissen \etal, 2000, Sen \etal, 2000),
the responses in cat auditory cortex to signals with 
naturalistic statistical properties (Rotman \etal, 
1999), and motion sensitive cells in the fly (Warzecha 
and Egelhaaf, 2001, de Ruyter van Steveninck \etal, 
2001). In each case compromises are struck between well 
controlled stimuli with understandable statistical 
properties and the fully natural case.

A more radical approach to natural stimulation was taken
by Roeder in the early sixties (see Roeder, 1998). He 
and his coworkers made recordings from moth auditory 
neurons in response to the cries of bats flying overhead
in the open field. More recently the visual system of 
{\it Limulus} was studied with the animal moving almost 
free on the sea floor (Passaglia \etal, 1997). 

Here we study motion sensitive visual neurons in the
fly, and---in the spirit of Roeder's work---rather than
trying to construct approximations to natural stimuli in
the laboratory, we take the experiment into nature. We
record the responses of H1, a wide field direction
selective neuron that responds to horizontal motion,
while the fly is being rotated along angular velocity
trajectories representative for free flying flies. These
trajectories indeed are quite wild, with velocities of
several thousand degrees per second and direction
changes which are complete within ten milliseconds. In
analyzing the responses to these stimuli we would like
to use methods that do not depend on detailed
assumptions about what features of the stimulus are
encoded or about what features of the spike sequences
carry this coded signal. Recently, information
theoretical methods were developed for analysing neural
responses to repeated sequences of otherwise arbitrarily
complex stimuli (de Ruyter van Steveninck \etal, 1997,
Strong \etal, 1998). In our experiments we repeat the
same motion trace, lasting several seconds, and this
provides us with the raw data for computing the relevant
information measures, as explained in section
\ref{sect:theory}. We emphasize that although we repeat 
the stimulus many times to estimate the relevant 
probability distributions of responses, the measures we 
derive from these distributions characterize the 
information coded by a single example of the neural 
response.

\section{Methods}

\subsection{Stimulus design considerations}

The giant motion sensitive interneurons in the fly's
lobula plate are sensitive primarily to such rigid
rotational motions of the fly as occur during flight
(Hausen, 1982, Krapp \etal, 1997), and these cells
typically have very large visual fields. It is this wide
field rigid rotation that we want to reproduce as we
construct a naturalistic stimulus. But what pattern of
rotational velocities should we use? As a benchmark we
will present data from an experiment where the fly was
rotated at velocities that remained constant for one
second each. We would, however, also like to present the
fly with stimuli that are more representative for
natural flight. Free flight trajectories were recorded
in the classic work of Land and Collett (1974), who
studied chasing behaviour in {\it Fannia canicularis}
and found body turning speeds of several thousand
degrees per second. A recent study (van Hateren and
Schilstra, 1999) reports flight measurements from {\it
Calliphora} at high temporal and spatial resolution. In
these experiments flies made of order 10 turns per
second, and turning speeds of the head reached values of
over 3000$^\circ$/s. In general, high angular velocities
pose problems for visual stimulus displays, because even
at relatively high frame rates they may give rise to
ghost images. In our laboratory we use a Tektronix 608
display monitor with a 500 Hz frame rate,. Then
3000$^\circ$/s corresponds to jumps from frame to frame
of 6$^\circ$, four times larger than the spacing between
photoreceptors. Although the frame rate used here is
well above the photoreceptor flicker fusion frequency
(de Ruyter van Steveninck and Laughlin, 1996) the
presence of ghost images may have consequences for the
encoding of motion signals. Further, the light intensity
of the typical displays used in the laboratory is much
lower than outside. As an example, the Tektronix 608
induces of order $5\cdot10^4$ photoconversions/s in fly
photoreceptors at its maximum brightness. The brightness
outside can easily be a factor of a hundred higher
(Land, 1981), although the photoreceptor pupil mechanism
will limit the maximum photon flux to about $10^6$
photoconversions/s (Howard \etal, 1987). Finally, the
field of view of H1 is very large, covering essentially
the field of one eye (Krapp and Hengstenberg, 1997),
which is about 6.85 sr or 55$\%$ of the full $4\pi$ sr
in female {\em Calliphora vicina} (estimates based on
Beersma \etal, 1977. See also Fig. 1). In practice with 
a display monitor it is hard to stimulate the fly with
coherent motion over such a large area and in most of
our laboratory experiments we stimulate less than about
20$\%$ of the full visual field of H1. 

\subsection{Stimulus apparatus}

All the factors mentioned above suggest an experimental
design in which the visual world can be made to move
more or less continuously relative to the fly, and this
is easiest to accomplish by moving the fly relative to
the world as occurs during free flight. We therefore 
constructed a light and compact assembly consisting
of a fly holder, electrode manipulator, and preamplifier
that can be mounted on a stepper motor, as shown in 
figure 1. This setup is rigid enough to allow high speed
rotations around the vertical axis while extracellular 
recordings are made from the H1 cell. Because it is 
powered by batteries the setup can be taken outside, so 
that the fly's visual system is stimulated with natural 
visual scenes. The mounting and recording stage 
inevitably covers some area in the fly's visual 
field. During the experiment this rotates along with the
fly, and so does not contribute to motion in the fly's 
visual field. By tracing the contours of the setup as 
seen from the fly, we estimate the shape and size of 
this overlap, as depicted in figure 1. The setup was 
designed to minimize the overlap in the visual field of 
the left eye. In the experiments presented here, 
recordings were therefore made from contralateral H1, on
the right side of the head. The setup occludes only 1.52
sr, or 22\%, of the visual field of the left eye, most 
of it ventral-caudal, as indicated by the heavy mesh in 
the right panel of figure 1.

The stepper motor (Berger-Lahr RDM 564/50) was driven by
a Divistep 331.1 controller in microstep mode, that is,
at $10^4$ steps per revolution, corresponding to a 
smallest step size of $0.036^\circ$ or roughly 1/30th 
of an interommatidial angle. The Divistep controller in 
turn was driven by pulses from a custom designed 
interface that produced pulse trains by reading pulse 
frequency values from the parallel port of a laptop 
computer. Pulse frequency values were refreshed every 2 
ms.

To generate naturalistic motion stimuli we used
published trajectories of chasing {\em Fannia} from 
Land and Collett (1974), interpolated smoothly between 
their 20 ms sample points. For technical reasons we had 
to limit the accelerations of the setup, and we chose 
therefore to rotate the fly at half the rotational 
velocities derived from the Land and Collett data. This 
may be reasonable as {\em Calliphora} is a larger fly 
than {\em Fannia}, and is likely to make slower turns. 
The constant velocity data presented in figure 2 were 
taken with rotation speeds ranging from about 
$0.28^\circ$/s to $4500^\circ$/s. To avoid extreme 
accelerations during high velocity presentations the 
pulse program for the stepper motor delivered smooth 100
ms pulse frequency ramps to switch between velocities. 
For velocities below $18^\circ$/s pulses were sent to 
the controller at intervals longer than 2 ms. At the 
lowest constant velocity used in our experiments, 
$0.28^\circ$/s, pulses were delivered at 128 ms 
intervals. The step size was small enough so that a 
modulation of the PSTH was undetectable in the 
experiment.

The experiment of figure 2 compares data from the
outdoor setup to data taken inside with the fly
observing a Tektronix 608 CRT. The stimulus displayed on
this monitor consisted of 190 vertical lines, with
intensities derived from a one-dimensional scan of the
scene viewed by the fly in the outdoor experiment. The
moving scene was generated by a digital signal
processor, and written at a 500 Hz frame rate. As
mentioned above, this gives rise to ghosting at high
image speeds when the pattern makes large jumps from
frame to frame. The DSP produced the coarse part of
motion essentially by stepping through lines in a buffer
memory. On top of this, fine displacements were 
produced by moving the entire image
by fractions of
a linewidth at each frame. The resulting 
motion was smooth and not limited to integer steps. The 
fly was positioned so that the screen subtended a 
rectangular area of $67^\circ$ horizontal by $55^\circ$ 
vertical, with the left eye facing the CRT and rightmost
vertical edge of the CRT approximately in the sagittal 
plane of the fly's head.

\subsection{Information theoretic analysis of neural 
firing patterns} \label{sect:theory}

We describe briefly a technique for quantifying 
information transmission by spike trains (de Ruyter van
Steveninck \etal, 1997, Strong \etal, 1998, de Ruyter 
van Steveninck \etal, 2001). 
We consider segments of the spike train 
with length $T$ divided in a number of bins of width 
$\Delta t$, where $\Delta t$ ranges from one millisecond
up to $\Delta t=T$. Each such bin may hold a number of 
spikes, but within a bin no distinction is made on where
the spikes appear. However, two windows of length $T$ 
that have different combinations of filled bins are 
counted as different firing patterns. Also, two windows 
in which the same bins are filled but with 
different count values, are distinguished. We 
refer to such firing patterns as words, $W_{T,\Delta t}$. From 
an experiment in which we repeat a reasonably long 
naturalistic stimulus a number of times, $N_r$ (here $N_r=200$ 
repetitions of a $T_r=5$ seconds long sequence) we get a large 
number of these words, $W_{T,\Delta t}(t)$, with $t$ the time 
since the start of the experiment. Here we discretize $t$ in 1 ms
bins, giving us 5000 words per repetition period, and 
$10^6$ words in the entire experiment. From this set of
words we set up word probability distributions, from 
which we calculate total and noise entropies, and their 
difference, according to Shannon's definitions: 

\begin{enumerate} \item The total entropy,
$S_{tot}(T,\Delta t)$. From the list of words
$W_{T,\Delta t}(t)$, for all $t$ 
($0\leq t \leq N_r\cdot T_r$), 
we directly get a distribution, $P(W_{T,\Delta
t})$ describing the probability of finding a word
anywhere in the entire experiment. The total entropy is
now: \begin{equation} S_{tot}(T,\Delta t)=-\sum_{W}
P(W_{T,\Delta t})\cdot \log_2[P(W_{T,\Delta t})]
\end{equation} This entropy measures the richness of the
`vocabulary' used by H1 under these experimental
conditions, hence the time of occurrence of the pattern
within the experiment is irrelevant. \item The average
noise entropy, $\bar{S}_{noise}(T,\Delta t)$. If the
neuron responded perfectly reproducibly to repeated
stimuli, then the information conveyed by the spike
train would equal the total entropy defined above. There
is noise, however, and this leads to variations in the
responses, as can be seen directly from the rasters in
Fig. 3. $\bar{S}_{noise}(T,\Delta t)$ gives us an
estimate of how variable the response to identical
stimuli is. We first accumulate, for each instant
$t_r$ in the stimulus
sequence, the distribution of all those firing patterns
$P(W_{T,\Delta t}|t_r)$, taken 
across all trials, that begin at $t_r$ (note that $0\leq
t_r \leq T_r$). The entropy of this distribution 
measures the (ir)reproducibility of the response at each
instant $t_r$: \begin{equation} S_{noise}(T,\Delta 
t,t_r)=-\sum_{W} P(W_{T,\Delta 
t}|t_r)\cdot\log_2[P(W_{T,\Delta t}|t_r)]. 
\end{equation} Calculating this for each point in time 
and averaging all these values we obtain the average 
noise entropy: \begin{equation} \bar{S}_{noise}(T,\Delta
t)=\frac{1}{T_r}\int_0^{T_r} S_{noise}(T,\Delta t,t_r) d
t_r. \end{equation} 
\item The information conveyed by words at the given length $T$
and resolution $\Delta t$ is the difference of these two
entropies: \begin{equation} I(T,\Delta t)=S_{tot}(T,\Delta
t)-\bar{S}_{noise}(T,\Delta t). \end{equation} The coding
efficiency of the spike train is the fraction of the total
entropy that is utilized to convey information: \begin{equation}
\eta(T,\Delta t)=\frac{I(T,\Delta t)}{S_{tot}(T,\Delta t)}.
\end{equation}
\end{enumerate} 

\noindent
Small values of $\eta(T,\Delta t)$ indicate a loose
coupling between stimulus and spike train, whereas
values close to 1 imply that there is little noise
entropy, so that most of the structure of the spike
train is meaningful, and carries a message. Here we will
not be interested in the decoding question, that is in
{\em what} that message is, but only in {\em how much}
information is conveyed about the stimulus. We will then
compare these values in different conditions. 

It should be stressed that the information values we
derive by these methods are not strictly {\em about}
velocity. They are potentially about anything in the
stimulus that is repetitive with period $T_r$. It is our
job as experimenters to construct inputs that we think
will stimulate the neuron well, and for H1, naturalistic
wide field motion seems to be a good choice. But that
does not necessarily mean that that is the best choice.
Further, the motion pattern is dynamic, and any
noiseless time invariant operation on this signal will
produce a result that has the same repeat period as the
original. Our information measures do not distinguish
these cases; specifically, our discussion is unaffected 
by the question of whether H1 encodes velocity, 
acceleration, or some nonlinear function of these 
variables. Questions of decoding are highly 
interesting, but at the same time difficult to tackle 
for stimuli of the type studied here, and we will leave 
them aside in this paper.

It is interesting to try and estimate $I(T,\Delta t)$ as
we let $T$ become very long, and $\Delta t$ very short,
as this limit is the average rate of information
transmission. Because calculating this limit requires
very large data sets, we focus here on the information 
transmitted in constant time windows, $T=30\,{\rm ms}$, 
as a function of $\Delta t$. We choose $T=30{\rm ms}$ 
because that amounts to the delay time with which a 
chasing fly follows turns of a leading fly during a 
chase (Land and Collett, 1974); the end result, that is 
the dependence of information transmission on $\Delta 
t$, was found not to depend critically on the choice of 
$T$.

To quantify noise entropy, the method described above
requires that a stimulus waveform be repeated. Although
it is possible in principle to quantify information
transmission based only on one repetition, using many
repetitions is easier in practice. In a sense this mode
of stimulation is still removed from the realistic
situation in which stimuli are not repeated at all.
Indeed, in our experiments there are hints that the fly
adapts to the stimulus somewhat over the first few
presentations of the 5 second long stimulus. The effects
of adaptation to dynamic stimuli are certainly 
interesting (Brenner \etal, 2000, Fairhall \etal, 2001),
but in the data we present here we skip the first few
presentations, and only analyze that part of the
experiment in which the fly seems fully adapted to the
ongoing dynamic stimulus. Inspection of 
the rasters in that phase shows no obvious trends, so 
that the fly seems to be close to stationary conditions.
In this regard our information measures are 
lower bounds, as deviations from stationarity will 
increase our estimate of the noise entropy, lowering 
information estimates.

\section{Results}

\subsection{Operating range for naturalistic motion stimuli}

In order to be sure that H1 receives no dominant motion
related signals from other modalities than vision we
rotated the fly either in darkness or under a cover that
turned along with the fly. This did not produce
discernible motion responses in H1. Strictly speaking
that does not exclude possible modulatory mechanosensory
input, which could be investigated in principle by
presenting conflicting visual and mechanosensory
stimuli. The possibility seems remote, however, and even
if true it would not invalidate our conclusions about
the dependence of H1's information transmission on
parameters of the visual stimulus. 

As a first comparison between laboratory and natural
conditions we present data from an experiment in which
H1 was excited by one second long episodes of motion at
constant velocity. These were presented at a range of
velocities from about 0.28$^{\circ}$/s to
4500$^{\circ}$/s. Outdoors the fly was placed in a
wooded environment and rotated on the stepper motor. In
the laboratory the same fly watched a vertical bar
pattern derived from a one dimensional scan of the
natural environment in which the outdoor experiment was
done. This pattern was displayed on a standard Tektronix
608 monitor, with a rectangular stimulated visual area
of 67$^{\circ}$ horizontal by 55$^{\circ}$ vertical. The
pattern moved at the same settings of angular velocity
as were used outdoors, but the indoor and outdoor
stimuli differed both in average light level and in
stimulated area. 

Figure 2 shows the average firing rates obtained from
the last half second of each velocity presentation. At
low velocities, up to about 20$^{\circ}$/s, the spike
rates for both conditions are not very different,
despite the large change in total motion signal present
in the photoreceptor array. Apparently the fly adapts
these differences away (see Brenner \etal, 2000). In 
both experiments the rate depends roughly 
logarithmically on velocity over an appreciable range 
and this is partly a result of adaptation as well (de 
Ruyter van Steveninck \etal, 1986). In the laboratory 
experiment the motion response peaks at about 
100$^{\circ}$/s, whereas in natural conditions the fly 
encodes velocities monotonically for an extra order of 
magnitude, its response peaking in the neighbourhood of
1000$^{\circ}$/s. This brings H1's encoding of motion
under natural conditions in the range of behaviourally
relevant velocities. A lack of sensitivity to high
speeds has been claimed both to be an essential result
of the computational strategy used by the fly, and to be
advantageous in optomotor course control (Warzecha and
Egelhaaf, 1998). These conclusions do not pertain to the
conditions in the outdoor experiment, where H1 responds
robustly and reliably to angular velocities of well over
1000$^\circ$/s. 

\subsection{Motion detection throughout the day}

Figure 3 shows spike train rasters generated by H1 in three
outdoor experiments, focusing on a short segment that illustrates
some qualitative points. Trace (a) shows the velocity waveform,
which was the same in all three cases. The experiments were
performed at noon (b), half an hour before sunset (c), and about
half an hour after sunset (d). Rough estimates of the photon flux
in a blowfly photoreceptor looking at zenith are shown beside the
panels. In all experiments the fly saw the same 
scene, with a spatial distribution of intensities 
ranging from about 5\% to 100\% of the zenith value. 

The figure reveals that some aspects of the response are
quite reproducible, and further that particular events in the 
stimulus can be associated reliably with small numbers of spikes.
More dramatically, the timing precision of the spike trains 
gradually decreases going from the noon experiment to the one 
after sunset. Higher photon rates imply a more reliable physical 
input to the visual system. The figure therefore strongly 
suggests that the fly's visual system utilizes this increased 
input reliability to compute and encode motion more accurately 
when the light intensity increases. This statement is 
ecologically relevant, as the conditions of the experiment 
correspond to naturally occurring light levels and 
approximately to the naturally stimulated visual area. 
To get a feeling for the spike timing precision in the 
three conditions we can simply look at the distribution 
of timing of the first spike generated after a fixed 
criterion time (for which we choose t=0.28 s in (b) and 
(c) and t=0.30 s in (d)). The jitter in the spike timing
across different trials has a standard deviation of 0.95
ms in (b), 1.4 ms in (c), and 5.8 ms in (d). The 
relative timing of spikes can be even more accurate: the
interval from the first to the second spike fired after 
the criterion time is 2.3$\pm$0.23 ms in (b), 
5.0$\pm$0.6 ms in (c), and 16$\pm$2.4 ms in (d). 
Compared to the rapid onsets and offsets of the spike 
activity at the higher photon fluxes, the stimulus 
varies rather smoothly, which means that the time 
definition of spikes with respect to the stimulus can be
much better than might be suggested by the stimulus 
bandwidth. An example can be seen in the rather smooth 
hump in the velocity waveform at about t=0.43 s, which 
induces on most trials a well defined response 
consisting of a sharply defined pair of spikes.

We quantify these impressions using the information
theoretic approach described briefly in section
\ref{sect:theory}. The result of this analysis is shown
in Fig. 4a-d, for the three different experiments
discussed above. Figure 4c clearly shows that the
information in a 30 ms window increases both when the
light intensity goes up, and when the spikes are timed
with higher accuracy. The increase in information with
increasing spike timing precision is most dramatic for
the highest light levels, indicating that coding by fine
spike timing becomes more prominent the better the input
signal to noise ratio. A comparison of figures 4a (total
entropy) and figure 4b (noise entropy) reveals that the
increase in information content with increasing light
levels is primarily due to an increase in total entropy:
The neuron's vocabulary increases in size as its input
becomes better defined. Figure 4d shows that at the two
highest light intensities the coding efficiency is of
order 0.5 at time resolution $\Delta t$=1 ms, 
increasing slightly for larger values of
$\Delta t$. In the darkest condition the efficiency 
decreases markedly for all values of time resolution. 

The right column of figure 4 compares experiments in which we
took data both outdoors and in the laboratory. These data are
from another fly, but the conditions of the outdoor experiment
were similar to those for the first fly at the highest light
level. After the outdoor experiment the fly was taken inside the
laboratory, and the same velocity stimulus as the one used
outside was repeated inside. In the laboratory, 
as before, the visual stimulus was presented on a 
Tektronix 608 CRT. Photoreceptors facing the 
CRT received about 5 $\cdot 10^4$ photons per 
second at maximum intensity, a value in between the 
light intensities seen by the first fly in the 
experiments just before and just after sunset (grey and 
black symbols in figure 4a-d). Two experiments were done
indoor, one in which the picture on the monitor 
consisted of vertical bars with a contrast pattern 
measured in a horizontal scan of the outdoor scene 
(filled triangles), the other a high contrast square 
wave pattern with contrast=1, and spatial 
wavelength=12.5$^\circ$ (filled squares). From figure 4g
we see that the information transmitted by H1 is much 
lower in the laboratory experiments than in the outdoor 
experiment, due to the smaller stimulated area and 
the lower light level. Figure 4e shows that the decrease
in information, as before, is mainly due to a lowering 
of the total entropy. The noise entropy also decreases 
(figure 4f), but not enough to compensate. Somewhat 
surprisingly, the experiment with the high contrast 
pattern indoors leads to a slightly higher coding 
efficiency than even the outdoor experiment.

\section{Discussion}

Outdoor illumination can easily be a hundred times
brighter than anything displayed on common laboratory
equipment, and in the outdoor experiment stimuli extend
over a large fraction of the fly's full visual field 
rather than being confined to a small flat monitor. Both
effects are relevant for our experiments, as the higher 
brightness leads to higher photoreceptor signal to noise
ratios (de Ruyter van Steveninck and Laughlin, 1996), 
and as H1's receptive field covers almost a hemisphere 
(Krapp and Hengstenberg, 1997). In moving from 
laboratory to outdoor conditions, both effects increase
the signal to noise ratio of the input available for 
computation of rigid wide field motion from the 
photoreceptor array. The question then is whether the 
fly's brain uses this improvement in input signal 
quality to produce more accurate estimates of visual 
motion, and/or increase its operating range of motion 
detection. Figure 2 shows that the range of velocities 
that are encoded increases markedly when the visual 
input becomes more reliable.

If the accuracy of information processing is limited by 
noise sources within the nervous system, we 
should observe a plateau, that is, information 
transmission should saturate at some defined level of 
input signal quality. There is some arbitrariness in the
choice of the level of input signal quality, however: In
principle we can surpass any degree of accuracy of the 
physical input signal by simply increasing the light 
intensity, and at some point the internal 
randomness of the brain's components must become 
the limiting factor in information processing. 
However, statements about the magnitude of internal 
versus external noise in sensory information processing 
are primarily meaningful in the context of 
reasonable, physiological levels of input 
signal quality. Those stimuli that the 
animal encounters naturally, taken at the high 
end of their dynamic range, would meet  
this criterion. For the case considered here the dynamic range 
refers to light intensity, size of stimulated visual field, and 
dynamics of motion. The data we recorded outdoors show no sign of
saturation in information transmission when the input signal 
quality increases. On the contrary, if we compare the rasters of 
figure 3b and 3c, we see that there is a marked improvement in 
the timing of spikes, even over the highest decade of light 
intensity ($2\times10^5$ to $3\times10^6$ photons/s at zenith per
photoreceptor). This improvement translates into a 
significant gain in information transmission, 
especially at fine time resolution, as 
shown in figure 4c. Thus, in computing motion from the 
array of photoreceptors, the fly's brain does not suffer
noticeably from information bottlenecks imposed by 
internal noise, under ecologically relevant conditions.

In our outdoor experiments, the information content of
the spike train varies primarily as a result of a
varying total entropy (figure 4a). The noise entropy
(figure 4b) appears to be almost constant as a function
of light level. One can distinguish two different ways
to increase information transmission through a channel.
The first is to encode the same messages 
more accurately, the second to increase the variety of
messages, keeping the accuracy of each individual
message the same. The first scheme implies constant
total entropy and decreasing noise entropy, the second
an increase in total entropy at constant noise entropy.
Our data suggest that as the visual input 
becomes more reliable, the fly chooses to increase the 
vocabulary of H1 to encode a wider variety of features
of the motion stimulus, keeping precision roughly 
constant.

\subsubsection*{Acknowledgments}

We thank Naama Brenner, Steve Strong and Roland Koberle for many 
pleasant and enlightening discussions.

\section*{References}

\begin{harvard}

\item[]
Beersma, D.G.M., Stavenga, D.G., and Kuiper, J.W. Retinal 
lattice, visual field and binocularities in flies. Dependence on 
species and sex. {\em J. Comp. Physiol.} {\bf 119,} 
207--220 (1977).

\item[] 
Brenner, N., Bialek, W., and de Ruyter van Steveninck, R.  Adaptive
rescaling maximizes information transmission, {\em Neuron}
26, 695--702 (2000). 

\item[] 
Dan, Y.,  Atick, J. J., and Reid, R. C.
Efficient coding of natural scenes in the lateral geniculate nucleus:
Experimental test of a computational theory,
{\em J. Neurosci.} {\bf 16,} 3351--3362 (1996).

\item[]
Fairhall, A.L., Lewen, G.D., Bialek, W., de Ruyter van 
Steveninck, R.R. Multiple timescales of adaptation in a 
neural code. To be published in {\em Advances In 
Neural Information Processing Systems 14}, 2001.

\item[]
Gallant, J. L., Conner, C. E., and van Essen, D  C.
Neural activity in areas
V1, V2 and V4 during free viewing of natural scenes compared to controlled
viewing  {\em NeuroReport} {\bf 9,} 1673--1678 (1998).

\item[] 
van Hateren, J. H.,and Schilstra, C.  Blowfly flight and optic flow II.
Head movements during flight, {\em J. Exp. Biol.}  {\bf 202,} 1491--1500
(1999).

\item[] 
Hausen, K. Motion sensitive
interneurons in the optomotor system of the fly. II.
The horizontal cells: Receptive field organization
and response characteristics. Biol. Cybern. {\bf 
46,} 67--79 (1982).

\item[]
Howard, J., Blakeslee, B., Laughlin, S.B. The 
intracellular pupil mechanism and the maintenance of 
photoreceptor signal to noise ratios in the blowfly 
{\em Lucilia cuprina}. Proc. R. Soc. Lond. B. {\bf 231,}
415--435 (1987).

\item[]
Krapp, H.G., and Hengstenberg, R. A fast stimulus 
procedure to determine local receptive field properties 
of motion-sensitive visual interneurons. Vision Research
{\bf 37,} 225--234 (1997).  

\item[]
Land, M. F. Optics and Vision in Invertebrates. 
In: Autrum, H. (ed) {\em Handbook of Sensory 
Physiology}. Springer, Berlin, Heidelberg, New York 
(1981), pp. 472--592.

\item[] 
Land, M. F., and Collett, T. S.  Chasing behavior of houseflies ({\it Fannia 
canicularis}). A description and analysis, {\em J  Comp Physiol} 
{\bf 89,} 331--357 (1974).

\item[] 
Passaglia C., Dodge F., Herzog E., Jackson S., Barlow R.
(1997): Deciphering a neural code for vision. Proc Natl
Acad Sci USA 94, 12649--12654.

\item[]
Rieke, F., Bodnar, D. and Bialek, W. Naturalistic 
stimuli increase the rate and efficiency of information 
transmission by primary auditory neurons. Proc. R. Soc. 
Lond. B {\bf 262,} 259--265.

\item[] 
Rieke, F., Warland, D., de Ruyter van
Steveninck, R. R., and Bialek, W.  {\em Spikes: Exploring the neural
code} (MIT  Press, Cambridge, 1997).
  
\item[] 
Roeder, K.D. (1998): Nerve cells and insect
behavior. Harvard University Press, Cambridge, MA.

\item[] 
Rotman, Y., Bar Yosef, O., and Nelken, I.
Responses of auditory--cortex neurons to structural
features of natural sounds,
{\em Nature} {\bf 397,} 154--157 (1999).

\item[] 
de Ruyter van Steveninck, R.R., Zaagman, W.H., 
Mastebroek, H.A.K. Adaptation of transient 
responses of a movement sensitive neuron in the visual 
system of the blowfly Calliphora erythrocephala. Biol. 
Cybern. {\bf 54,} 223--236 (1986).
 
\item[] 
de Ruyter van Steveninck, R.R., Laughlin, S.B.:
Light adaptation and reliability in blowfly
photoreceptors. Int. J. Neural. Syst. {\bf 7,} 437--444.
(1996).  

\item[] 
de Ruyter van Steveninck, R. R., Lewen,  G. D.,
Strong,  S. P., Koberle, R., and Bialek, W. Reproducibility and 
variability in neural spike trains. {\em Science} {\bf 275,} 1805--1808 
(1997).

\item[]
de Ruyter van Steveninck, R., Borst, A., Bialek, W. 
real-time encoding of motion: Answerable questions and 
questionable answers from the fly's visual system. 
In: J.M. Zanker and J. Zeil (eds.) Motion vision. 
Computational and ecological constraints. Springer, 
Berlin, Heidelberg, New York (2001), pp. 279--306.

\item[] 
Sen, K., Wright, B. D., Bialek, W., and Doupe, 
A. J. Discovering features of natural stimuli relevant 
to forebrain auditory neurons. {\em Soc. Neurosci. 
Abstract} {\bf 551.2} (2000).

\item[] 
Stanley, G. B., Li, F. F., and  Dan, Y.
Reconstruction of natural scenes from ensemble responses in the lateral
geniculate nucleus,
{\em J. Neurosci.} {\bf 19,} 8036--8042 (1999).

\item[] 
Strong, S. P., Koberle, R., de
Ruyter van Steveninck, R. R., and Bialek,  W.  Entropy and
information in neural spike  trains. {\em Phys Rev Lett} {\bf 80,}
197--200 (1998).

\item[] 
Theunissen, F. E., and  Doupe, A. J.
Temporal and spectral sensitivity of complex auditory neurons in the
nucleus HVc of male zebra finches,
{\em J. Neurosci.} {\bf 18,} 3786--3802 (1998).

\item[] 
Theunissen, F. E.,  Sen, K., and   Doupe, A. J.
Spectral--temporal receptive fields of nonlinear auditory neurons obtained
using natural sounds,
{\em J. Neurosci.} {\bf 20,} 2315--2331 (2000).

\item[] 
Vickers, N.J., Christensen, T.A., Baker, T. , 
and Hildebrand, J.G. Odour-plume dynamics influence the 
brain's olfactory code. {\em Nature} {\bf 410,} 466--470 (2001).

\item[] 
Vinje, W. E., and  Gallant, J. L.
Sparse coding and decorrelation in
primary visual cortex during natural vision,
{\em Science} {\bf 287,} 1273--1276 (2000).

\item[] 
Warzecha, A.--K., and Egelhaaf, M. On the
performance of biological movement detectors and ideal velocity  sensors
in the context of optomotor course stabilization, {\em Vis. Neurosci.}
{\bf 15,} 113--122  (1998).

\item[]
Warzecha, A.--K., and Egelhaaf, M. Neural encoding of 
visual motion in real-time. 
In: J.M. Zanker and J. Zeil (eds.) Motion vision. 
Computational and ecological constraints. Springer, 
Berlin, Heidelberg, New York (2001), pp. 239--278.

\end{harvard}

\newpage

\section*{Figure legends}

\begin{figure}[ht]
\caption{
Left: Setup used in the outdoor experiments. The fly is in a
plastic tube, head protruding, and immobilized with wax.
A small feeding table is made, from which
the fly can drink sugarwater. The part of the setup 
shown here rotates around the axis indicated at the 
bottom, by means of a stepper motor. A silver reference
wire makes electrical contact with the body fluid, while
a tungsten microelectrode records action potentials 
extracellularly from H1, a wide-field motion sensitive 
neuron in the fly's lobula plate. The electrode signals 
are preamplified by a Burr-Brown INA111 integrated 
instrumentation amplifier, the output of which is fed 
through a slip ring system to a second stage amplifier 
and filter and digitized by a National Instruments 
PCMCIA data acquisition card in a laptop computer. The 
part of the setup visible in the figure is mounted on a 
stepper motor which is driven by computer-controlled 
laboratory built electronics. Right: Occlusion in the 
left visual field of the fly. The dot in centre 
represents the position of the fly.
The animal is looking in the direction of the arrow and 
has the same orientation as the fly in the setup at 
left. The thin mesh bordered by the heavy line 
represents the excluded part of the visual field of the 
left eye for a free flying fly (based on Beersma et al. 
1977). The heavy mesh represents the overlap of the left
eye's natural visual field with those parts of the setup
that rotate along with the fly, and therefore do not 
contribute to a motion signal. The total visual field of
the left eye is 6.85 sr, or $0.55 \cdot 4\pi$. The 
overlap depicted by the heavy mesh subtends about 1.52 
sr, or 22\% of the visual field of the left eye. } 
\end{figure}

\begin{figure}[ht]
\caption{
A comparison of responses to constant velocity
in a typical laboratory experiment (closed squares), and
in an outdoor setting where the fly is rotating (open
circles). Average firing rates were computed over the 
last 0.5 seconds of a 1 second constant velocity 
presentation.
} 
\end{figure} 

\begin{figure}[ht] \caption{ Responses of
the H1 neuron to the same motion trace recorded outside
at different times of the day. {\bf(a)} Short 
segment of the motion trace executed by the 
stepper motor with the fly. The full segment of 
motion lasted 5 seconds, and was derived from video 
recordings of natural fly flight during a chase (see Methods)
{\bf(b)} 50 Spike rasters in 
response to the motion trace in {\bf(a)}, taken at 
noon. {\bf(c)} As {\bf(b)}, but recorded about half an 
hour before sunset. {\bf(d)} As {\bf(b)}, but recorded 
about half an hour after sunset.} 
\end{figure}

\begin{figure}[ht]
\caption{Lefthand column: Information theoretic 
quantities for the three outdoor experiments whose 
rasters are shown in figure 3. The symbol 
shadings refer to the different conditions of 
illumination in the experiments. All figures refer to a 
30 ms measurement window in which neural firing 
patterns are defined at time resolutions, $\Delta t$, of
1,~2,~3,~5,~10,~15, and 30 ms, as given by the 
abscissae. {\bf(a)}: Total entropy of spike firing 
patterns. {\bf(b)}: Average noise entropy. {\bf(c)}: 
Average information transmitted by firing patterns. 
{\bf(d)}: Coding efficiency, defined as the transmitted
information divided by the total entropy.

Righthand column: The same quantities as plotted in the 
lefthand column, but now for an experiment outdoors 
(open symbols), and two experiments in the laboratory 
(closed symbols, see text for further description of 
conditions). Squares are for a moving square wave 
pattern of high contrast (C=1), and spatial wavelength 
12.5$^\circ$; triangles are for a moving sample of 
the natural scene at the location where the outdoor 
experiments were done. Both these stimulus patterns 
were generated on the cathode ray tube in the 
laboratory. }
 \end{figure} 

\end{document}